\documentclass[10pt,preprint]{aastex}

\font\smcap=cmcsc10

\begin{document}

\title{Keck Spectroscopy of Red Giant Stars in the Vicinity of M31's Massive
Globular Cluster G1\footnote{Data presented herein were obtained at the W.\
M.\ Keck Observatory, which is operated as a scientific partnership among the
California Institute of Technology, the University of California and the
National Aeronautics and Space Administration.  The Observatory was made
possible by the generous financial support of the W.\ M.\ Keck Foundation.}}

\author{David B.\ Reitzel}
\affil{Department of Physics \& Astronomy, 8371 Mathematical Sciences
Building, University of California at Los Angeles, Los Angeles, California
90095, USA\\
\email{\tt reitzel@astro.ucla.edu}}

\author{Puragra Guhathakurta}
\affil{UCO/Lick Observatory, University of California at Santa Cruz, 1156
High Street, Santa Cruz, California 95064, USA\\
\email{\tt raja@ucolick.org}}

\vskip 0.5 truecm
\and

\author{R.\ Michael Rich}
\affil{Department of Physics \& Astronomy, 8371 Mathematical Sciences
Building, University of California at Los Angeles, Los Angeles, California
90095, USA\\
\email{rmr@astro.ucla.edu}}

\begin{abstract}

We present results from an ongoing Keck spectroscopic survey of red giant
stars in a field located along the major axis of M31, $\approx34$~kpc in
projection from the nucleus and near the luminous globular cluster G1.  We
use multislit LRIS spectroscopy to measure the Ca\,{\smcap ii} near-infrared
triplet in 41~stars ranging in apparent magnitude from $20<I<22$.  Of these,
23~stars are found to have radial velocities $v<-200$~km~s$^{-1}$ indicating
that they are giants in M31; the rest are likely to be foreground Galactic
dwarf stars.  Roughly two-thirds of the M31 members concentrate at
$v=-451$~km~s$^{-1}$, with a relatively small velocity spread
[$\rm\sigma(Gaussian)=27~km~s^{-1}$] which suggests that they belong to the
outer disk or possibly a cold debris trail in the halo.  The mean velocity of
this group of red giants is consistent with that of nearby neutral hydrogen
and models of the velocity field of M31's disk, rather than with G1 or the
systemic velocity of M31.  We use $V,I$ photometry to estimate a mean
metallicity of $\rm\langle[Fe/H]_{phot}\rangle=-0.8$ for this group of
potential M31 outer disk stars.  Six stars out of the 23 M31 member giants
have metallicities and velocities consistent with those of G1 (after
accounting for its intrinsic spread in $v$ and [Fe/H]): one of these stars
lies within the projected tidal radius of G1 and is a likely member; the
remaining 5~stars are not physically close to G1, and may represent tidal
debris from G1.  However, more data are needed to confirm the nature of these
5~stars, as it is likely that they simply represent M31's field halo
population.  We might have expected to detect tidal debris if G1 were the
remnant core of a dwarf galaxy being accreted by M31; instead, the majority
of M31 giants in this field are metal-rich and belong to what is evidently
the outer disk of M31, and only a small fraction ($\lesssim20\%$) could
possibly have originated in G1.

\end{abstract}


\section{Introduction}

The outer regions of M31 have become an increasingly complicated field of
study as it has become clear in recent years that the role of accretion in
halo formation is of considerable importance.  \citet{fer02} present
star-count maps showing what appear to be extensive tidal disturbances in the
halo of M31, including in the vicinity of the massive globular cluster G1.
\citet[][hereafter RG02]{rei02} find evidence for a subtle stream-like
feature, in an outer halo field located 19~kpc from the center in projection
along the southeastern minor-axis, using a combination of kinematics and
metallicity measurements of red giant branch (RGB) stars.  \citet{guh02}
confirm that this feature continues along the minor axis in two inner halo
fields located near the globular clusters G312 and G302, located 11 and 7~kpc
from the nucleus of M31 respectively near the southeastern minor axis.  M31's
two closest satellites M32 and NGC~205 are known to be undergoing tidal
stripping \citep*{cho02}; yet there is no definite proposal for a companion
that might have been responsible for the large-scale streams seen in the
halo.

The area around G1 is a particularly interesting field to study, as this
object has been proposed to be the core of a tidally-disrupted dwarf galaxy
\citep{mey01}.  If this is the case, one might expect to find the tidal
debris surrounding the main body of the object with velocities and
metallicites similar to G1 itself.  In addition, the field around G1 is
expected to have roughly equal numbers of M31 halo and disk stars
\citep*{rei98,hod95} so this gives us the opportunity to study the disk
population of M31 further out ($r\sim34$~kpc) than has been done to date.
\citet{fer01} study a field 30~kpc from the nucleus of M31 along the NE major
axis and estimate a mean metallicity of $\rm[Fe/H]\simeq-0.7$; they find that
the population is mostly old ($>8$~Gyr) but there is evidence for an
intermediate-age population as well.  In a paper that accompanies this work,
\citet{ric03} report evidence for an intermediate-age population (6--8~Gyr)
only 2~kpc from the location of G1.  \citet{dav03} examines the outer disks
of NGC~2403 and M33 and finds evidence for an intermediate-age population
well outside the current star-forming disk in both galaxies.

The observations (imaging and spectroscopy), data reduction, and analysis are
presented in \S\,2, a discussion follows in \S\,3, and a summary in \S\,4.

\section{Observations and Data Analysis}

\subsection{Imaging/Stellar Photometry}

The Low Resolution Imaging Spectrograph \citep[LRIS;][]{oke95} on the Keck~II
10-m telescope is used in both imaging and spectroscopic modes for this
project.  The LRIS imaging data were obtained on August~15, 1998 (UT) and
consist of single 5-min exposures in the $V$ and $I$ bands centered on the
globular cluster G1.  The seeing FWHM was about $0.8''$ and the sky
conditions photometric.  The $2048\times1600$~pixel images cover
$7.3'\times5.7'$ with a pixel scale of $0.215''$.  After standard initial
processing of the data---2D overscan subtraction, trimming, flat fielding,
and fringe removal (in the $I$-band only), the DAOPHOT software package
\citep{ste87} is used for object detection and $(V,~I)$ photometry.
Star-galaxy separation is done on the basis of image morphology.  Several
astrometric standard stars from the USNO Catalog happen to lie on our LRIS
images and this allows for accurate calibration of the stellar ($x$,~$y$)
positions into ($\alpha$,~$\delta$) coordinates.

Archival {\it Hubble Space Telescope\/}\footnote{Based on observations made
with the NASA/ESA Hubble Space Telescope obtained at the Space Telescope
Science Institute, which is operated by the Association of Universities for
Research in Astronomy, Inc., under NASA Contract NAS~5-2655.} ({\it
HST\/})/Wide Field Planetary Camera~2 (WFPC2) images from programs GO-5464
and GO-5907, two pointings centered on G1 but at different telescope roll
angles, are also analyzed.  Each pointing consists of images in the F555W and
F814W bands (roughly equivalent to the $V$ and $I$ bands, respectively), with
total exposure times per band of 1660 and 1280~s (GO-5464) and 2200 and
1800~s (GO-5907).  The WFPC2 images are processed through the standard {\it
HST\/} pipeline, and the HSTphot package \citep{dol00} is used to obtain a
list of stellar positions and ($V$,~$I$) photometry on the Johnson/Cousins
system based on the transformation relations of \citet{hol95}.

The agreement in stellar magnitudes between the two WFPC2 pointings is better
than 0.1~mag, and the measurements are averaged for stars common to both
pointings.  The WFPC2 pointings are contained entirely within the LRIS
imaging field and cover only its central portion.  The LRIS $V$ and $I$
magnitude zeropoints are adjusted to match WFPC2 photometry using stars in
the overlap region; the rms difference between WFPC2 and (adjusted) LRIS
magnitudes is 0.1~mag.  Stellar photometry with WFPC2 is somewhat more
accurate than with LRIS, so the former is used wherever possible ($\sim30\%$
of the final sample of stars).

\subsection{Spectroscopy}

A sample of 57~spectroscopic targets in the apparent magnitude range
$20<I<22$ were selected from the above stellar photometry/astrometry list.
Two multi-slit masks were designed using Drew Phillips' {\tt simulator}
software, one having 30~slitlets and the other 27~slitlets on M31 RGB star
candidates, with the targets distributed more-or-less uniformly over the LRIS
field of view.  Multi-slit spectroscopic observations were carried using
Keck/LRIS during a two-night run on September~28--29, 1998 (UT).  Each
spectrum covers the spectral range
$\lambda\lambda7550\>$--$\>8850\,\rm\AA$
containing the near-infrared Ca\,{\smcap ii} triplet: $\lambda\lambda8498$,
8542, and 8662\,\AA.  The instrumental spectral/velocity resolution is
1.94\,\AA/68~km~s$^{-1}$ (FWHM); see RG02 for details.  The total exposure
times for the two masks is 1.7 and 2.0~hr.  Individual exposures are
typically 30~min long, although some exposures were stopped short due to
telescope, instrument, and weather problems.

Overscan correction, 2D bias structure subtraction, and cosmic ray removal
are accomplished using standard IRAF\footnote{IRAF is distributed by the
National Optical Astronomy Observatories, which are operated by the
Association of Universities for Research in Astronomy, Inc., under
cooperative agreement with the National Science Foundation.} tasks.  Cosmic
rays are removed on the basis of object sharpness and peak pixel brightness.
They are masked from each image along with a surrounding 1-pixel buffer to
remove the low-level wings of each event.  A flat-field correction for each
data frame is performed using a spectral dome flat that is well-matched to
the data frame in terms of LRIS flexure effects.  Data reduction
issues/complications for LRIS spectra are discussed in some detail in RG02;
that study did not however use the \citet{phi03} reduction software that is
used here.

Wavelength calibration, sky subtraction and extraction are all accomplished
using the LRIS data reduction pipeline developed by \citet{phi03}.  The
software uses an optical model for the various elements of the LRIS
spectrograph (collimator, grating, camera, etc.) to derive a mapping from the
slitmask to the CCD detector as a function of wavelength.  This optical model
is based on spectrograph design drawings, and has been empirically refined
using calibration spectra (arc lamp through a grid-of-holes mask) taken close
to the observing run to account for misalignment of any of LRIS' elements.
Even so the model is only good to about 0.3~pixels ($0.06''$ in the spatial
direction and $\sim$0.2\AA\ in the dispersion direction).  In addition, one
must allow for small time-dependent alignment/focus errors, and variations
from exposure to exposure because of instrument flexure.  These errors are
removed by low-order corrections: a zeropoint correction in wavelength
measured from a bright night-sky emission line, and a plate-scale and offset
which are solved for using the measured locii of slitlet edges.

\subsection{Radial Velocity Measurement}

In order to determine the radial velocity of each object, its final coadded
spectrum is cross-correlated against a template spectrum (average spectrum of
three~control sample stars from RG02, where the coaddition is done after
shifting each to zero velocity).  The cross-correlation function (CCF) is
computed from $-1000$ to +1000~km~s$^{-1}$, covering a plausible range of
radial velocities for stars associated with M31.  The CCF technique yields an
unambiguous peak and a reliable radial velocity for 41 of the 57~objects
comprising the main sample of potential M31 targets.  A complete description
of this CCF procedure is given in RG02.

The radial velocity determined from the location of the CCF peak, $v_{\rm
obs}$, is corrected to the heliocentric frame using the task {\smcap RVCOR}
in IRAF.  We estimate the rms error in radial velocity from the degree of
significance of the CCF peak: $\sigma_v=\sigma_v^{\rm TD}(1+r_{\rm TD})^{-1}$
\citep{ton79}.  The value of $\sigma_v^{\rm TD}$ is empirically found to be
77~km~s$^{-1}$ for our instrumental set-up (RG02).  The mean $1\sigma$ error
in velocity for this G1 field sample is 21~km~s$^{-1}$.

The success rate for radial velocity measurements in our sample is 41 out of
57, or about 72\%.  This is somewhat lower than the 80\% success rate in the
RG02 study, but the difference can be attributed to two~factors:
\begin{itemize}
\item[1.]{RG02 estimate that less than half their failures, about 7\% of the
spectroscopic sample, are background field galaxies.  The surface density of
M31 giants is probably higher near G1 ($r\sim34$~kpc, major axis) than in
RG02's field ($r\sim19$~kpc, minor axis): similar numbers of M31 halo stars
given the 5:3 apparent flattening of the halo \citep{fer02}, but a larger
number of M31 disk stars \citep{hod95}.  This would nominally imply a {\it
lower\/} galaxy contamination fraction in our G1 field assuming an isotropic
galaxy distribution.  However, the RG02 spectroscopic sample was pre-screened
against galaxies using $UBRI$ photometry \citep{rei98}, in addition to
standard morphological star-galaxy separation, whereas no such color
selection is done in the present study.  Thus, a higher fraction of galaxies
may have slipped into our G1 field spectroscopic sample.}
\item[2.]{The remainder of the failures in the RG02 study are thought to be
a result of inadequate S/N.  The total exposure times for our G1 field
spectroscopic masks are a factor of 2--3 shorter than for the RG02 masks.
Based on tests using subsets of the RG02 data, we expect the failure rate due
to inadequate S/N to be about 20\% for the G1 field sample presented here.}
\end{itemize}

\subsection{Metallicity Estimation}

\begin{figure}
\plotone{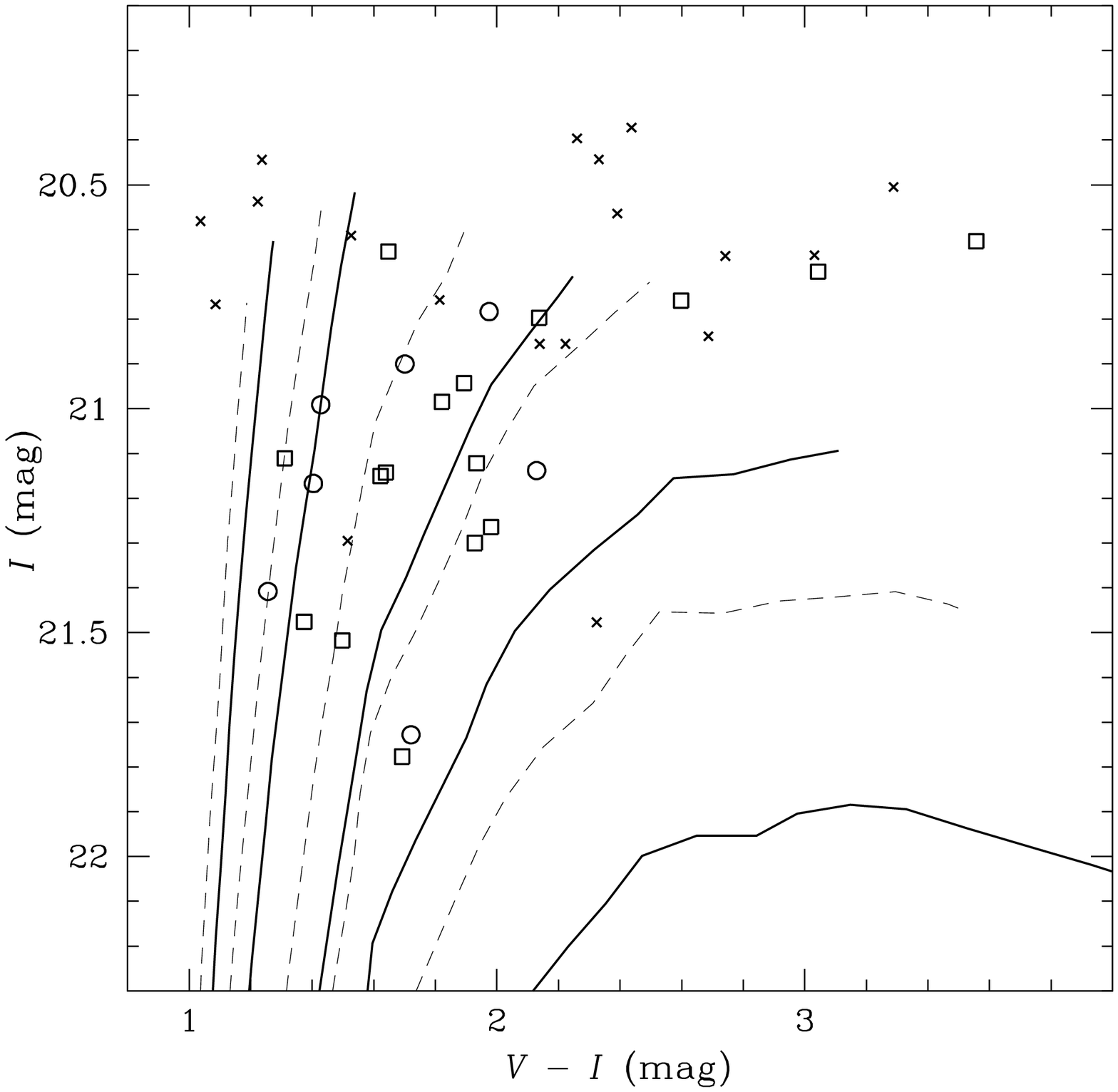}
\figcaption[Reitzel.fig01.eps]{\label{vi_cmd}{Color-magnitude diagram
    of the spectroscopic targets 
    based on $V$ and $I$ photometry from LRIS and {\it HST\/}/WFPC2 images.
    Only the 41~targets for which a radial velocity measurement was possible
    are shown: $v>-200$~km~s$^{-1}$ (``$\times$''; likely foreground Milky
    Way dwarf stars); $-400<v<-200$~km~s$^{-1}$ (open circles);
    $v<-400$~km~s$^{-1}$ (open squares; similar kinematics to the disk of
    M31).  Isochrones from the Padova group \citep{gir00} with
    $\rm[Fe/H]=-2.3$, $-1.3$, $-0.7$, $-0.4$, and $-0.02$
    (left$\rightarrow$right) and $t=12.6$~Gyr and $t=4.0$~Gyr are shown as
    solid and dotted lines, respectively.  Note, younger isochrones yield
    somewhat higher metallicity estimates for the stars.}}
\end{figure}

The metallicity of each star is estimated from its position in the
($V-I$,~$I$) color-magnitude diagram Fig.~\ref{vi_cmd}, by comparing it to
model isochrones from the Padova group \citep{gir00} with $\rm[Fe/H]=-2.3$,
$-1.7$, $-1.3$, $-0.7$, $-0.4$, $-0.02$, and $+0.18$ and $t=12.6$~Gyr.
The isochrones are translated into apparent/observed $I$ vs.\ $V-I$ space
based on an adopted M31 distance of 783~kpc \citep{sta98,hol98}, or a true
distance modulus of $(m-M)_0=24.47$, and a mean reddening of
$\langle{E(B-V)}\rangle=0.06$ towards G1 derived from the \citet*{sch98} dust
map.  A standard slope of $R_V=3.1$ is assumed for the Galactic dust
extinction law, which translates into $E(V-I)/E(B-V)=1.4$ \citep*{car89}.
The resulting model isochrones are fit with a Legendre polynomial of 6th
order in $V-I$ and 10th order in $I$ to interpolate between the isochrones.
This yields a photometric estimate of each star's metallicity,
$\rm[Fe/H]_{phot}$.  If the actual age of the stellar population is closer to
4~Gyr instead of the value of 12.6~Gyr adopted in this paper \citep{ric03},
the metallicity estimates would be revised upwards by about $+0.3$~dex (see
Fig.~\ref{vi_cmd}).

The errors in metallicity are dominated by systematics in the method.  The
relative metallicities of the sample may be ranked to within 0.1~dex, but the
true metallicity of each star is uncertain by at least 0.25~dex due to
systematic errors such as differential reddening, age error/spread,
variations in the degree of alpha enhancement, and inaccuracies in the
models.  The systematic error of 0.25~dex is added in quadrature to a random
error component of 0.1~dex: conservatively, our metallicities have overall
errors on the order of 0.27~dex.

The above error estimates apply only to stars located within the range of the
model isochrones in Fig.~\ref{vi_cmd}.  It is clear from the CMD though that
a significant number of stars lie above the tip of the RGB and/or are bluer
than the most metal-poor isochrone.  As discussed in \S\,3, most of these
outliers are foreground Galactic dwarf stars for which the $\rm[Fe/H]_{phot}$
estimate is in any case meaningless.  Two of the outliers are probable
members of M31's disk (see below); we caution that their extrapolated
$\rm[Fe/H]_{phot}$ estimates are very uncertain.

RG02 derive spectroscopic metallicity estimates from Ca\,{\smcap ii} line
strengths.  The lower S/N of our spectra makes such estimates unreliable so
they are not presented here.  We have verified though that the gross features
of the [Fe/H] vs.\ radial velocity plot [Fig.~\ref{feh_vel}$\>$(bottom)]
remain unchanged when the photometric metallicity estimates are replaced with
spectroscopic ones.

\section{Discussion}

\begin{figure}
\plotone{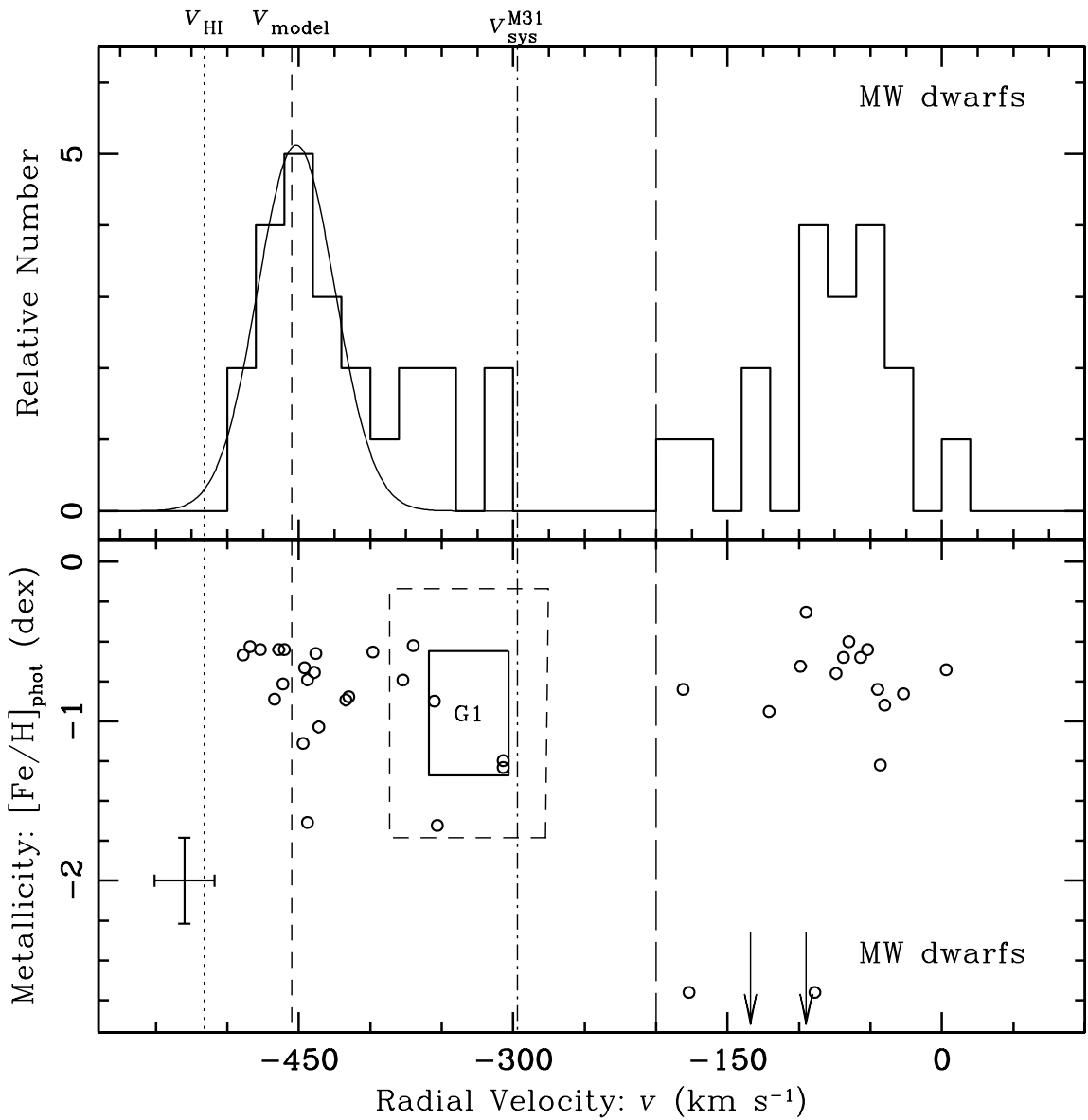}
\figcaption[Reitzel.fig02.eps]{\label{feh_vel}{({\it
    Top\/})~Distribution of heliocentric radial 
    velocities for the main sample.  The peak near $v\sim-75$~km~s$^{-1}$
    represents primarily Milky Way dwarf stars.  The concentration near
    $v=-450$~km~s$^{-1}$ is consistent with the velocity of M31's HI disk
    $10'$ east (dotted line) and especially with the model of M31's velocity
    field along this specific line of sight \citep[short-dashed
    line,][]{saw81}, but not with the globular cluster G1 (see lower panel)
    or M31's systemic velocity (dot-dashed line).  Thus, these stars likely
    represent the M31 disk population and not tidal debris from G1.~~~  ({\it
    Bottom\/})~Radial velocity versus iron abundance, the latter estimated
    from each star's position in the ($V-I$,~$I$) color-magnitude diagram
    (Fig.~\ref{vi_cmd}).  Stars with $v>-200$~km~s$^{-1}$ (long-dashed line)
    are likely to be Milky Way dwarf stars; the illustrated values of
    $\rm[Fe/H]_{phot}$ for these objects are meaningless as the metallicity
    estimation method is only appropriate for M31 red giants.  The
    $\pm1\sigma$ measurement errors in radial velocity and metallicity are
    shown in the lower left.  The solid and dashed rectangles indicate G1's
    1-$\sigma$ and 2-$\sigma$ (respectively) internal spread in metallicity
    and velocity \citep{mey01}.  Six objects lie within $\pm2\sigma$ of G1's
    values; one is a likely a member of the globular cluster as its projected
    position lies within G1's tidal radius.  The remaining five objects may
    represent tidal debris from G1 or may belong to M31's smooth halo
    population; the majority of metal-rich stars in the field appear to
    belong to the disk of M31, not its halo.}}
\end{figure}

The distribution of heliocentric radial velocities in our sample shows a
clear concentration near $v\simeq-450$~km~s$^{-1}$ and another near
$v\simeq-75$~km~s$^{-1}$ [Fig.~\ref{feh_vel}$\>$(top)].  Based on this
observed distribution, the sample can be crudely divided into three main
components: Milky Way dwarf stars, M31 halo giants, and M31 disk giants.  We
expect that most of the 18~objects with $v\ge-200$~km~s$^{-1}$ are foreground
Galactic dwarf stars \citep[RG02;][]{rat85} and do not consider these for any
further analysis.

Seven objects lie in the velocity range $-400<v<-200$~km~s$^{-1}$.  They are
likely to be M31 field halo RGB stars, with some possibly representing tidal
debris from G1.  One star is potentially a member of the G1 globular cluster:
it lies $30''$ from the cluster center, well within its tidal radius of
$54''$ \citep{mey01} in projection; its metallicity $\rm[Fe/H]_{phot}=-0.74$
is within the $\pm1\sigma$ intrinsic [Fe/H] spread (0.39~dex) around G1's
mean metallicity of -0.95~dex \citep{mey01}; its radial velocity
$v=-377$~km~s$^{-1}$ is within the $\pm2\sigma$ intrinsic velocity dispersion
($\sigma_v=28$~km~s$^{-1}$) around G1's systemic velocity of
$v=-331$~km~s$^{-1}$ \citep{mey01}.  Three of the remaining ``halo-like''
objects have velocities and metallicities which are within $\pm1\sigma$ of
G1's values and two more lie within $\pm2\sigma$
[Fig.~\ref{feh_vel}$\>$(bottom)].  However, none of these five~stars are
physically close to G1, the closest being more than two tidal radii away.
These stars may represent tidal debris from G1; however the small size of our
sample makes it impossible for us to draw any firm conclusions.

The remaining 16~objects with $v<-400$~km~s$^{-1}$ appear to be tightly
clustered around a mean velocity of $-451$~km~s$^{-1}$.  This is similar to
the velocity of the 21-cm neutral hydrogen line, $v_{\rm
HI}\sim-516$~km~s$^{-1}$ for the gaseous disk $10'$ due east of G1; however,
no HI gas has been detected at the exact location of G1 \citep{thi03}.
\citet{saw81} fit a linear density-wave model \citep*{lin69} to the mean
rotation curve derived from the high-sensitivity HI survey by \citet*{cra80}
in order to construct a velocity field map of M31.  Their model predicts a
disk velocity of $v\simeq-455$~km~s$^{-1}$ at our field location (G1 is
at a projected distance of $r\sim34$~kpc from the galaxy center near its SW
major axis).  \citet{sof81} use the mean rotation curve observed for M31 and
assume an inclination of $i=77^{\circ}$ to construct a radial velocity field:
this predicts a disk velocity of $v\simeq-480$~km~s$^{-1}$ near G1.  Both
predictions agree with our measured mean velocity to within the errors.  A
Gaussian centered at $v=-451$~km~s$^{-1}$ with $\sigma=27$~km~s$^{-1}$ fits
the distribution well [Fig.~\ref{feh_vel}$\>$(top)].  This observed velocity
spread is roughly equivalent to (if slightly larger than) our estimated
velocity measurement error.  Subtracting the measurement error in quadrature
from the measured dispersion yields an {\it intrinsic\/} dispersion of less
than 20~km~s$^{-1}$ for the sample of ``disk-like'' RGB stars, although this
value is not well constrained because of the small sample size.  The rms
velocity dispersion of the HI gas is measured to be 8.1~km~s$^{-1}$
independent of position within M31's disk \citep{unw83}.  \citet{nol87} find
a velocity dispersion of about 38~km~s$^{-1}$ for a sample of M31 ``disk''
planetary nebulae located somewhat closer to the galaxy's center
($15<r<30$~kpc), but it is possible that some of these objects belong to the
dynamically hotter halo or thick disk components.

These ``disk-like'' RGB stars near G1 have a mean metallicity of
$\rm\langle[Fe/H]_{phot}\rangle_{\rm disk}=-0.8$.  The [Fe/H] distribution
appears to be somewhat asymmetric: the peak is shifted towards the metal-rich
end ($-0.55$~dex) with an extended tail towards metal-poor values
[Fig.~\ref{feh_vel}$\>$(bottom)].  By contrast, the seven ``halo-like'' stars
have a mean metallicity of $\rm\langle[Fe/H]_{phot}\rangle_{\rm halo}=-0.99$.
It should be noted, however, that the metal-rich end of these distributions
are not well constrained because the spectroscopic sample is
magnitude-limited (RG02): the tip of the RGB is not as bright for a
metal-rich population as it is for a metal-poor one and this leads to a bias
against metal-rich stars.  In addition, our demarcation between ``halo'' and
``disk'' samples at $-400$~km~s$^{-1}$ is arbitrary: it is possible that a
small fraction of our ``disk-like'' stars actually belong to M31's halo.
 
Our results on M31 outer disk stars can be compared to two other studies
targeting the outer disks of spiral galaxies.  \citet{dav03} finds mean
metallicities of $\rm\langle[Fe/H]\rangle\lesssim-1$ in the outer parts of
the Sc galaxies M33 and NGC~2403.  \citet*{fer98} find gas phase [O/H]
abundances that are roughly 10\% solar in the far outer regions
(1.5$\>$--2$\>R_{25}$) of late-type disk galaxies; assuming a solar [O/Fe]
ratio, this gas would form stars with $\rm[Fe/H]\sim-1$.  The spiral galaxies
in these studies are less luminous and of later Hubble type than M31, so it
is perhaps not surprising that M31's outer disk is more metal rich than
theirs.  In addition, the accompanying paper by \citet{ric03} describes
stellar photometry in a deep {\it Hubble Space Telescope\/}/Wide Field
Planetary Camera~2 field located only 2~kpc in projection from G1, and finds
evidence for a population of objects with an age of 6--8~Gyr.  Two of the
disk-like stars in our spectrscopic sample lie well above the tip of the RGB
(Fig.~\ref{vi_cmd}) and could be representative of an intermediate-age
asymptotic giant branch population.  This is consistent with the findings of
the \citeauthor{dav03} study.  Thus, the cumulative evidence leads us to
believe that the majority of metal-rich stars in our sample represents a
population that formed in a disk.  It is also possible, though less likely,
that this population instead represents a merger event that happens to have
the same radial velocity as the outer disk of M31.

\section{Summary}

We use Keck/LRIS multi-slit spectroscopy to measure the Ca near-infrared
triplet in 41~candidate M31 giants in the apparent magnitude range $20<I<22$.
We find 23~stars have radial velocities that are consistent with membership
in M31; two-thirds of these M31 member candidates likely belong to a disk
population.  The disk stars have a mean heliocentric radial velocity of
$v=-451$~km~s$^{-1}$.  A Gaussian centered at this velocity with
$\sigma=27$~km~s$^{-1}$ fits the distribution indicating a true velocity
dispersion in the disk of $\sigma_{\rm intrinsic}<20$~km~s$^{-1}$ (after
accounting for velocity measurement error).  This ``disk-like'' population
has a mean metallicity of $\rm[Fe/H]=-0.8$; most of the stars are slightly
more metal-rich than this, with a tail to lower metallicities.  Although the
velocities and metallicities are consistent with a disk population, we cannot
rule out the possibility that these stars belong to a dynamically-cold stream
from a recent satellite merger event.

Six stars have metallicities and velocities that are roughly consistent with
G1's values.  One of these stars lies within the tidal radius of G1 (at least
in projection) and is thus likely to be a member of the globular cluster.
The remaining five could possibly represent tidal debris from G1, but are
more likely to simply be M31 field halo stars.  Even if they do represent G1
tidal debris, they make up no more than $20\%$ of the overall population, as
most of the metal-rich stars in the field appear to belong to M31's disk.  A
larger sample is needed in order to determine the exact contributions of
these various populations to M31's outer disk region.

\acknowledgments

Support for proposal GO-9099 was provided by NASA through a grant from the
Space Telescope Science Institute, which is operated by the Association of
Universities for Research in Astronomy under NASA contract NAS~5-26555.
P.G.\ acknowledges support from the National Research Council of Canada in
the form of a 2002-2003 Herzberg fellowship, and is grateful to the Herzberg
Institute of Astrophysics for hosting him during that time.  He would also
like to thank Joe Miller for his support through a UCO/Lick Observatory
Director's grant.  We thank Marla Geha for a careful reading of the
manuscript.  The authors wish to recognize and acknowledge the very
significant cultural role and reverence that the summit of Mauna Kea has
always had within the indigenous Hawaiian community.  We are most fortunate
to have the opportunity to conduct observations from this mountain.

\clearpage

\clearpage
\end{document}